\documentclass[a4paper,twoside]{article}

\usepackage{epsfig}
\usepackage{subcaption}
\usepackage{calc}
\usepackage{amssymb}
\usepackage{amstext}
\usepackage{amsmath}
\usepackage{amsthm}
\usepackage{multicol}
\usepackage{pslatex}
\usepackage{apalike}
\usepackage[bottom]{footmisc}
\usepackage{graphicx}
\graphicspath{ {./figures/} }
\usepackage{xcolor}
\usepackage{booktabs}
\usepackage{SCITEPRESS}     

\begin{document}

\title{Students’ interests related to web and mobile technologies study}

\author{\authorname{Manuela Andreea Petrescu\sup{1}\orcidAuthor{0000-0002-9537-1466}, and Adrian Sterca\sup{1}\orcidAuthor{0000-0002-5911-0269}, and Ioan Badarinza\sup{1}\orcidAuthor{0000-0001-8233-8264}}
\affiliation{\sup{1}Department of computer Science, Babes-Bolyai University, Str. M. Kogalniceanu, Cluj-Napoca, Romania}
\email{{manuela.petrescu, adrian.sterca, ioan.badarinza}@ubbcluj.ro}
}



\abstract{We explore in this paper the interests and challenges of students regarding web and mobile technologies. Our study is based on a survey among undergraduate students, students that attend a Web Programming course. In particular, we study the challenges students have in following a successful career in web or mobile development and we have found that the most important one is the large effort required for keeping up to date with the fast changing web and mobile technologies. Overall, the attitude of the surveyed undergraduate students towards web development and mobile development is rather positive, as more than 60\% of them said that they are interested in a career in web or mobile development. We also found out that most of them prefer working on back-end web technologies. As for the specific web technologies students are interested on, they are highly varied. Overall, our study provides valuable insights into the interests and challenges of students regarding web and mobile technologies, which can guide the development of effective teaching and learning approaches in this area.}

\onecolumn \maketitle \normalsize \setcounter{footnote}{0} \vfill

\section{\uppercase{Introduction}}

Web and mobile development are essential components of modern business, as they enable companies to connect with their customers through digital platforms. In today's world, the global population stands at 7.91 billion \cite{digital22}, and is expected to reach 8 billion by mid-2023, given the annual growth rate of 1.0 percent. More than half of it, specifically 57.00\% of the world's population resides in urban regions according to the same report. The technology got into urban areas, as into the most remote and inaccessible locations. As a result, even the populations in underdeveloped countries have access to technology so the number of persons that are using the internet grew. According to \cite{digital22}, in the first months of 2022, the number of global internet users has risen to 4.95 billion, which is equivalent to 62.5\% of the world's population. it is important to notice the sharp internet penetration trend, as data indicates that there has been a 4.0\% increase, or 192 million new users, in the past year. Some of the new users connected to the internet using specific devices (laptops, computers, tablets), and mobile phones; mobile phone usage increased sharply globally. More than two-thirds of the global population, 67.1\% of the world’s population now uses a mobile phone. The mobile phone adoption trend is increasing because technology becomes more accessible in terms of price and performance. Over the past year, the total number of mobile users has increased by 1.8\%, in fact, there are 95 million new users compared to the same period last year \cite{digital22}.

Web and mobile development are essential components of modern business, as they use the infrastructure and huge numbers of potential clients, thus enabling companies to connect with their customers through digital platforms. In today's world, where most people are using smartphones and other digital devices, web, and mobile presence are essential to reach a broader audience. The market trends demand for designing and creating websites for the internet is increasing. Mobile development for applications for mobile devices is also on an ascending trend. Both web and mobile development play a crucial role in promoting products, services, and brands online, enabling businesses to increase their target population by increasing their online presence. Web and mobile can provide a personalized experience for users, using advanced technologies such as AI and machine learning, and can find and adjust to the specific needs and preferences of their customers. Subsequently, companies can target their customers with specific advertising campaigns, increasing their sales. This helps to build a loyal customer base and improve customer satisfaction, which ultimately leads to increased revenue and profits.

According to market trends, web and mobile development are crucial components of modern business as more and more potential clients can be found online, especially from mobile. Mobile usage overtook other devices, as in 2020, 55\% of website traffic came from mobile devices \cite{Statista}. The size of information passed through the internet grew from around 50 exabytes (1 exabyte equals 1 billion gigabytes) to more than 300 exabytes in 2022 according to the global internet traffic reports made by Statista \cite{Statista}. The traffic was mainly generated by different applications usage, applications that offer a range of benefits, including personalized experiences for customers, valuable data insights, and provide a competitive advantage. The companies will continue to invest in developing web and mobile applications, providing a demand for specialized workforce programmers to develop the applications.

With this setup in mind, we wanted to find out if (1) the second year university students in computer science are interested to have a career path in web or mobile development and which technologies would they like to work with, (2) what are the challenges they are confronted with, and (3) what are the reasons for which they choose to work or not in these fields.

The structure of this paper is as follows: Section~\ref{sec:literature_review} examines recent studies on related topics,  Section~\ref{sec:methodology} provides a comprehensive overview of the methodology used for the experiment for data collection and interpretation, as well as the set of participants. Section~\ref{sec:data_collection} describes the data collection and analysis of the research questions. Sections ~\ref{sec:student_challenges}, ~\ref{sec:student_interests_career} and ~\ref{sec:student_interests_technology} present our findings after analyzing the collected responses to our research questions. Section ~\ref{sec:student_challenges} describes the most important challenges students have in following a career path in web or mobile development, Section ~\ref{sec:student_interests_career} analyzes the students' interests in such a career and Section ~\ref{sec:student_interests_technology} discusses their preferred technologies in these fields. 
Possible threats to validity are discussed in Section~\ref{sec:threats_to_validity}. Finally, Section~\ref{sec:conclusion} presents the concluding remarks.

\section{\uppercase{Literature review}}
\label{sec:literature_review}

The industry of building web and mobile applications is growing every day and it has a huge demand for skilled developers. Teaching students how to develop web applications is a very important part of their study years. There were multiple attempts done by professors to structure the curriculum in such a way so that the students understand the variety of technologies and environments for web development. For example, in \cite{jevremovic2018wide}, the authors present a centralized and collaborative approach to teaching Web development. During their study, they noticed that students have a lot of issues understanding and troubleshooting the environments used for web applications and they have tried to build a centralized solution for web development as a teaching process. They noticed that students had a better understanding of the web environments after this change. 

Many studies recommend adding team project work to university courses as requirements for students in order to better prepare the students for the following stage in their career as a junior software developer in a company \cite{iacob2019}, \cite{heberle2018}. 

Still, one of the biggest challenges that computer science students are facing when learning how to build a web application is the constantly changing technologies \cite{park2011learning}. Despite these challenges that students have learning the technologies, recent studies have shown that graduates are well-prepared for the job market. They still need to go through a training process at the beginning after they are hired, but this is more for understanding the company's culture and the way that their employees work \cite{lundberg2021employable}.

Previous studies have shown that IT industry is interested in a few skills when they are hiring new employees. Among these skills, problem-solving skill is the most important one \cite{radermacher2014investigating}. The other very important ones are communication and teamwork \cite{clokie2016graduate}, \cite{begel2008novice}.

\section{\uppercase{Methodology and Setup}}
\label{sec:methodology}
To find out the response to the paper's research questions, we performed an online survey. The participant set was formed by second-year computer science students enrolled in the Web Programming course. The main purpose of the Web Programming course is to teach the students how to build a web application using front-end and back-end technologies. It has an extended syllabus including more than a dozen languages.
On the front-end side, the course includes information about HTML and XML languages for the web (e.g. XHTML, XSLT), CSS, Javascript, Javascript libraries like jQuery and some Javascript frameworks like Angular, and on the back-end side, it provides foundations on how to use PHP, Java, and .Net to build the back-end of an application. The students have to work on a lot of assignments to get some experience in all of these technologies so that they can fully understand what they like and what path they want to pursue. The course also focuses on popular design patterns for web applications like the Model-View-Controller and popular architectures for web applications like the REST API architecture. Another focus of the course if on the security of web applications which is ever so challenging nowadays. So the Web Programming course is quite abundant in technologies. it is also one of the most dynamic courses as the web technologies change at a very rapid pace. 

As a research method, we used the survey research method, defined in ACM Sigsoft Empirical Standards for Software Engineering Research \cite{ACM}. We had open questions, to allow students to express themselves and to be able to get a clearer in-site into their interests. We also had closed questions, to be able to check exactly (and measure with accuracy) some aspects of their stated interests. We paid attention to the order of the questions in the survey, for example, we asked them first what technologies are they interested in (the last questions were not visible on the page), and only after they completed the answer and get to the last questions we provided them with a list of technologies to choose from (to mark which technology are they particularly interested in). We chose this approach to find out what are they interested in without influencing their answers, and only at the end of the survey after they mentioned their technology interests, we suggested other possible options.

We send the survey link to the students during the laboratory and asked them to complete it at their own will, during or after the laboratory. Around 47\% of the responses were completed after the lab was over. When we sent the link, we also mentioned the anonymous and optional character of the survey as the purpose of collecting the data.  

\subsection{Participants}
The target participant set that took part in this survey was formed by second-year university students enrolled in the Web Programming course; the students were aleatory allocated to have laboratories with two of the authors of the study. The allocation process was formed by two steps, in the first one, the students were alphabetically sorted after the last name, and based on this list, they were allocated to a group. In the second step, the groups were aleatory allocated to the authors of this study by third parties (when the timetable was created).
The total number of students that were asked to participate in the study and got the survey link was 85. The number of answers was 61, we considered it was a good percentage that allows us to have valid results, results that correctly reflect their perceptions, ideas, and interests. When deciding to pursue analyzing the responses, we validate their number against the number of answers in other published studies \cite{petrescu23,motogna21} in computer science (with less or a similar number of participants). We concluded that the participant set is large enough and representative to have a valid study.

\section{\uppercase{Data Collection}}
\label{sec:data_collection}
Data was collected using Google Forms. We opt for this tool due to its usability, friendliness, and because it was most probably already used by students. Students trust the tool to provide anonymity. For us, it was useful as we could see responses in real-time. The survey remained open for ten days. This period allowed us to send it to the students from different groups (that had classes at different times during the week); so they would have a three days grace period after the last laboratory to answer the survey (if they wanted to do it). 

For this research, we used quantitative methods, and more specifically questionnaire surveys as they were defined by community standards \cite{ACM}, and for interpreting the test we used thematic analysis as defined by \cite{Braun19}. This research survey method, where collected data was interpreted using thematic analysis, was previously used in other computer science-related papers \cite{motogna21,redmond13}. The standard mentioned a procedure that implies steps that should be followed, so one author performed the following activities when analyzing the data:
\begin{itemize}
    \item collected the data
    \item performed a brief analysis of the open answers with the purpose to reallocate them to other survey questions (if they were better fitted elsewhere)
    \item determined the keywords and grouped them into classes
    \item the other author verified the keywords classification, 
    \item the other authors performed some observations
    \item the observations were analyzed by all authors and the decisions were incorporated into the results.
\end{itemize}
The opened answers sometimes contained more than one keyword, some answers contained one keyword and we had questions for which we did not get an answer. Due to this aspect, the number of identified keywords is larger than the number of answers. To have a clear idea about the key items' prevalence, the results are expressed as a percentage between the number of key items and the number of answers. Due to this fact, the sum of percentages will exceed 100\%. 
The questions asked in the survey can be visualized in Table \ref{tab:questions}.
\begin{table}[h!]
  \caption{Survey Questions}
  \label{tab:questions}
  {\small{
  \begin{tabular}{p{7.0cm}}
\toprule
Q1. Would you like to pursue a career in web development? (Choose from Yes or No options)\\
Q2. Please mention your reasons. \\
Q3. What technologies/topics do you want to learn?\\
Q4. Which technology do you think is best in a long run (if you choose between web and mobile)?\\
Q5. Which are in your opinion the main challenges for a web/mobile software developer?\\
Q6. What type of web technologies do you want to pursue in a future career: front-end technologies or back-end  technologies? (Choose front-end or back-end options) \\
Q7. Which technologies are you interested in? Multiple choice of PHP, JS, React, Ajax, jsp/servlets, Angular, .NET, Client-server architecture, Site performance, Site reliability, Site security, Python flask, node.js, Java Springboot, Real-time communication over web (WebRTC), Micro-service architecture for web applications, REST API web servers.\\
\toprule
\end{tabular}
}}
\end{table}

\section{\uppercase{Students' challenges in a career in web / mobile development}}
\label{sec:student_challenges}

\begin{figure}[!ht]
\centering
\includegraphics[width=0.5\textwidth]{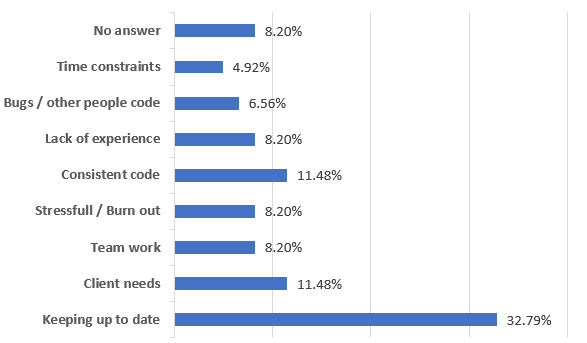}
\caption{Reported challenges} 
\label{fig:challenges}
\end{figure}

By far, as can be visualized in Figure \ref{fig:challenges}, the highest challenge reported by students was the rapid change of technologies and the need to adapt to them. 

The largest group of students 32.79\% reported this aspect: \textit{''Constantly adapting to newer and newer versions of everything''}, or \textit{''Keeping up with the evolving technology''}. It seems that even for the younger generation, perpetual adaptation is perceived as a stress factor. The next reported challenge is tightly connected with the industry: client needs/ issues \textit{''Understanding what the client needs.'', ''Dealing with customers''}, or \textit{''Agreeing with the client on the project''}. Writing clean, consistent code and maintaining the quality of delivered code are other challenges reported by a rather large group of students: 11.48\%. There are mentioned challenges tightly related to soft skills: teamwork and communication appear in 11.47\% of the total answers: \textit{''working on a team and communication''}. Other two challenges are related: stressful/burn-out mentioned by 8.20\% and time constraints mentioned by 4.92\%: \textit{''Time pressure. "Checklist" things (for example - do 5 tickets per week)'', ''Communication and the fight to show results but do not burn out.'', ''Overcoming the burnout''}. Students seem to be confronted with a lack of experience, and 8.20\% of them perceive it as a challenge: \textit{''Having experience''}, they also mentioned the difficulty to work on \textit{''other's people code''}. The documentation proved to be a difficult to use tool, some students mentioned that is hard to look for a solution in thick and too overwhelming documentation. 
Other emerging technologies are perceived as direct competitors, a fight between men and machines, between programmers and AI. AI is perceived as a competitor by 3.27\% of them: \textit{''Also - at this moment, a huge challenge especially for juniors are the developments of AI technologies (for example, at this moment, ChatGPT is better than me)''}. 

In conclusion, the main challenge is the fact that they have to keep up with rapidly changing technologies, and the soft skills challenges represent an important part (even if it is for communication, teamwork, or understanding client needs). Time constraints and burnout are other main challenges. AI concurrence, learning process, lack of documentation, or too thick documentation are other challenges mentioned by students.

\section{\uppercase{Students' interest in a career in web / mobile development}}
\label{sec:student_interests_career}

To find out the students' interest in following a career path in web or mobile development, we analyzed the answers provided to three questions: the stated intention (first question), their reasons (answers from the second question), and if they prefer working on front-end side or back-end side of web applications.  

\begin{figure}[!ht]
\centering
\includegraphics[width=0.5\textwidth]{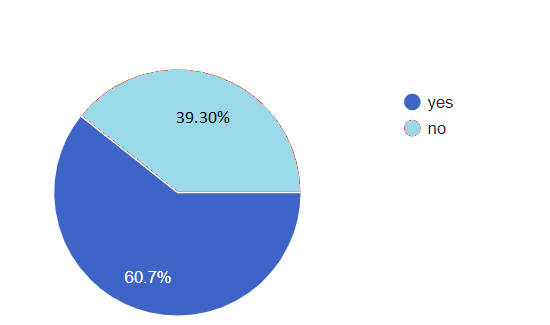}
\caption{Stated web-related work intention} 
\label{fig:work}
\end{figure}

More than half of the students (around 60.7\% from the participant set) mentioned that they would like to have a career path in web or mobile development as can be visualized in Figure \ref{fig:work}.

\begin{figure}[!ht]
\centering
\includegraphics[width=0.5\textwidth]{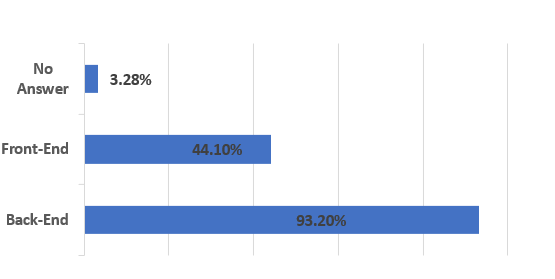}
\caption{Students' preferences} 
\label{fig:preferences}
\end{figure}

When we asked the students if they prefer working with front-end technologies or with back-end, we allowed them to select one option, both options, or neither one of them. Almost all the students (93.20\%) would like to work in the back-end, around half of them (44.20\%) mentioned they would like to work in the front-end and 3.94\% did not answer this question - Figure \ref{fig:preferences}. Based on their answers, the students do not want to work in front-end in a large percentage, some of the reasons being detailed as answers to the second question: \textit{''I do not like frontend''} or \textit{''Not particularly in web development because you need to pay attention to small stuff like the proper margin/alignment, I prefer to make things work (back-end).''}. 

\textbf{Students' reasons for choosing/not choosing web or mobile development} 
\\
\\
The most prevalent reason for choosing a web/mobile related career path is related to its trendiness 16.39\%, web technologies, and their usage are perceived as worth learning because they represent: \textit{''A continuing growing industry''} and \textit{''Everyone wants a website''}. Students are well connected to the market trends, their replies suggest a positive correlation between their working intention in a specific field and market demand.
The interest in technology (14.74\%) and passion (6.55\%) are other reasons that attract students to web/mobile development: \textit{''I guess it's cool'', ''Yes, but I'm not sure, for a moment I am a mobile developer but I had an Internship on web dev which was very engaging''}. In the front-end back-end comparison, students prefer to work on the back-end (their preference is enforced by detailed explanations), and the back-end appreciations are positive: \textit{''I like back-end more
I don't like the front-end part of it, but the back-end is alright''}. 9.83\% of the students mentioned specifically that they  like front-end \textit{''I don't like frontend''} and other 11.47\% stated they like back-end: \textit{''You need to pay attention to small stuff like the proper margin/alignment, I prefer to make things work (back-end).''}. As opinions are divided, there was a percent of 9.83\% of students who clearly stated they like front-end: \textit{''I grew way more fond of front-end development than back-end because I find it more entertaining '', ''I like colors, shapes, and animations and designing things'', ''I like designing pages and it doesn't seem that stressful''}.
Other students just wanted something else, and they specified it: AI was the most mentioned option. \textit{''Other areas of programming are more interesting.'', ''I am more interested in different topics such as AI'', ''AI is more interesting to me.''}

In conclusion, students are relatively interested in a career path in web or mobile development, but they would prefer to work mainly in the back end. Front-end is perceived to require a lot of attention to detail and to be ease replaceable by new technologies such as AI. Some students prefer the front end because it is more entertaining, it implies page design, and is perceived as less stressful. 

\subsection{Job taxonomy/market demand}

To gain insights into the job market, we look into studies referred to computer science literature and technical reports, including \cite{RePEc22}, and \cite{ESCoE18}, which evaluated job requirements based on job postings. A web developer or mobile developer job position is usually at the top when it comes to the market needs.

The US Bureau of Labor Statistics prospects that the market demand on web developers in the US is projected to increase by 23\% in the 2021-2031 period, much faster than the average increase of all other occupations.

A general taxonomy of career paths in web development would be the following:
\begin{itemize}
    \item front-end developer
    \item back-end developer
    \item full-stack developer
    \item UX designer
    \item web designer
    \item mobile web developer
    \item DevOps engineer
    \item security engineer
\end{itemize}

But each of the above career paths can further be specialized into more specific paths, especially the first three of them (e.g. a back-end web developer can be a Python web developer, a Java web developer, a C\# web developer etc.).


\section{\uppercase{Students' interest in technologies}}
\label{sec:student_interests_technology}

\begin{figure}[!htbp]
\centering
\includegraphics[width=0.48\textwidth]{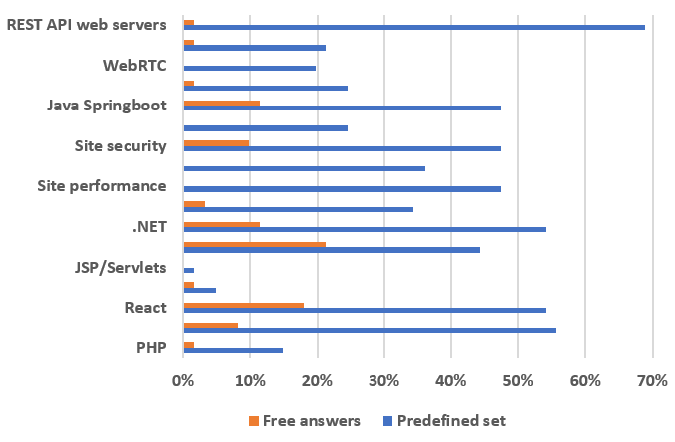}
\caption{Students' interest in technologies} 
\label{fig:interest}
\end{figure}

To find out what is student's genuine interest in technologies, but also to take into account their current knowledge, we used the answers provided by students to the questions \textit{''What type of web technologies do you want to pursue in a future career'',  ''Which technologies are you interested in'', and, ''Which technology do you think is best in a long run''}. The last question was the simplest one from this set, 37.70\% of the students mentioned that mobile is the best in a long run, as the greater percent 62.29\% appreciated that web is best in long run. All the students answered this question. However, when we asked them to write which technologies are they interested in, the results were variate, and the percentages quite different compared to the answers when they had to choose a technology from a predefined set. For example, for React, only 18.03\% mentioned it compared to 54.10\% of students that choose it from the predefined set. We consider for the web course we should take into account the predefined set, but their answers (no predefined answers) are more accurate to find out their overall interests in computer science technologies. It was also interesting to notice that when students were asked to mention themselves technologies, 9.83\% did not answer, but in the case when the technologies list was predefined, all of them answered by selecting more than one technology. A comparison of the percentages for each technology is displayed in Figure \ref{fig:interest}.

Except for the technologies mentioned above, when asked to answer by themselves, a very small percentage of students' answers contained specific technologies \textit{''Kafka'', ''Google services'', ''Vue.js'', and ''Java''}. A larger group of students, 13.11\% provided generic answers: \textit{''Web''} without specifying exactly what technology, \textit{''Backend''} was mentioned by 9.83\%, and \textit{''Game Development''} by 6.55\%. \textit{''Server development''} appeared only in 3.27\% of the answers, as \textit{'mobile''} in 1.63\%. \textit{''AI''} was mentioned by a larger percentage, 16.39\%, but in this case, we got generic answers like ''AI'' and very specific answers \textit{''AI, Inverse Kinematics, procedural generation, marching cubes algorithm, Perlin noise'', ''Deep learning, machine learning''}. Some students do not want a specific technology as long as it could provide good growing options: \textit{''Whatever is used in the present in the industry.''}.

In conclusion, the students can be grouped into two categories: well prepared and documented students, which know exactly what they want to study, and a group of students that want to study but prefer to be told what to learn (they selected a lot of predefined options). In general, students have a large interest in AI; regarding web/mobile related technologies they are interested in REST APIs, .Net, Java SpringBoot, React, JS, Angular, site performance, and security.

\section{\uppercase{Threats to Validity}}
\label{sec:threats_to_validity}
During the elaboration of our study, we carefully considered the guidelines and recommendations for survey research outlined in \cite{ACM}. We identified several potential threats to validity that we tried to mitigate the risk associated with them. We paid attention to the target population and participant selection, drop-out rates, authors' subjective approach, and research ethics.
To address the concerns related to the target population and participant set, we sent the survey to all the second-year, computer science students enrolled in the web course that were randomly allocated to specific teachers (authors of this paper). Due to this aspect, we did not engage in any participant selection,  as everyone was automatically involved. All students were given the opportunity to participate if they so chose. The target participant set was randomly allocated, and the teachers did not have any involvement in the allocation process. We also took steps to mitigate drop-out rates by limiting the survey to one online questionnaire and keeping the number of questions manageable to enable participants to complete it quickly. We also considered the participant set size, and compared to the size from other computer science published studies \cite{motogna21,gerster2020}, our conclusion was that we have a sufficient number of participants to generate valid results.
We recognized the possible bias that the students responded positive, and in order to mitigate the risks we announced them that all the responses are anonymous, validate them (we had answers stating they they do not like WEB, or want to work in AI). We also compared to similar study (same university, similar participant set - second years students from computer science aleatory selected), in which students stated that are not interested in database and SQL related jobs \cite{pop23} (congruent with our results). Based in these facts, we concluded that the students replied honestly regarding their intentions.

We recognized the potential for subjectivity in our approach and we mitigated this risk by following recommended research procedures and guidelines for text interpretation. To ensure ethical research practices, we informed students that participation was voluntary, and we made it clear about the anonymous nature of the survey. We also provided information on the purpose of the data collection and how we intended to use the information gathered.

\section{\uppercase{Conclusion and future work}}
\label{sec:conclusion}
This study was performed with the purpose to find out if second-year computer science students are interested in a career in web or mobile development, in technologies, and what are their main challenges. The participant set was randomly selected, the research method was a survey questionnaire and we applied thematic analysis to interpret the open questions. We took into consideration the scientific guidelines for this type of research mentioned in \cite{ACM}, and we took measures to eliminate possible author subjectivity. We found out that students are interested in a career in web or mobile, even if when they were asked to specify exactly which technologies are they interested in, the larger set of technologies was not web or mobile-related. 
However, when students were provided with a fixed set of technologies, they selected much more technologies compared to the questions when they were asked to answer by themselves. This proves that they were not exactly sure, but are willing to learn. The main reason for pursuing a web/mobile related career was related to the ''trendiness'' and perceived market requirements \textit{''Everyone wants a website''}. This aspect is correlated to the fact that the second option related to career path (except web) is AI, another technology perceived to be ''in trend''. Specifically, some answers mentioned the financial aspects and the fact that students expect to find well-paid jobs in this domain. In terms of working on the back-end or front-end, the larger majority preferred back-end.

Even if it was not the purpose of this study, we found out that the students that already work (having part-time or full-time jobs) stated they feel stressed, they are fighting with time and pressure to deliver and they experience burnout or they fear they would get in that state. Moreover, they feel it is hard to cope with the rapid changes in the computer science domain, with changing technologies.

Other studies should be performed to find out the degree of correlation between the ''trendlines'' of a domain with the student's stated intention to have a career in that domain.

\bibliographystyle{apalike}

\begin{thebibliography}{}

\bibitem[Begel and Simon, 2008]{begel2008novice}
Begel, A. and Simon, B. (2008).
\newblock Novice software developers, all over again.
\newblock In {\em Proceedings of the fourth international workshop on computing
  education research}, pages 3--14.

\bibitem[Braun et~al., 2019]{Braun19}
Braun, V., Clarke, V., Hayfield, N., and Terry, G. (2019).
\newblock {\em Thematic Analysis}, pages 843--860.
\newblock Springer Singapore.

\bibitem[Clokie and Fourie, 2016]{clokie2016graduate}
Clokie, T.~L. and Fourie, E. (2016).
\newblock Graduate employability and communication competence: Are
  undergraduates taught relevant skills?
\newblock {\em Business and Professional Communication Quarterly},
  79(4):442--463.

\bibitem[Djumalieva1 and Sleeman, 2018]{ESCoE18}
Djumalieva1, J. and Sleeman, C. (2018).
\newblock An open and data-driven taxonomy of skills extracted from online job
  adverts.
\newblock Economic Statistics Centre of Excellence (ESCoE) Discussion Papers
  ESCoE DP-2018-13, Economic Statistics Centre of Excellence (ESCoE).

\bibitem[Gallagher et~al., 2022]{RePEc22}
Gallagher, E., Kerle, I., Sleeman, C., and Richardson, G. (2022).
\newblock A new approach to building a skills taxonomy.
\newblock Economic Statistics Centre of Excellence (ESCoE) Technical Reports
  ESCOE-TR-16, Economic Statistics Centre of Excellence (ESCoE).

\bibitem[Gerster et~al., 2020]{gerster2020}
Gerster, D., Dremel, C., Brenner, W., and Kelker, P. (2020).
\newblock How enterprises adopt agile forms of organizational design: a
  multiple-case study.
\newblock {\em ACM SIGMIS Database: the DATABASE for Advances in Information
  Systems}, 51(1):84--103.

\bibitem[Heberle et~al., 2018]{heberle2018}
Heberle, A., Neumann, R., Stengel, I., and Regier, S. (2018).
\newblock Teaching agile principles and software engineering concepts through
  real-life projects.
\newblock In {\em Proceedings of the 2018 IEEE Global Engineering Education
  Conference (EDUCON)}, pages 1723--1728.

\bibitem[Iacob and Faily, 2019]{iacob2019}
Iacob, C. and Faily, S. (2019).
\newblock Exploring the gap between the student expectations and the reality of
  teamwork in undergraduate software engineering group projects.
\newblock {\em Journal of Systems and Software}, 157:1--18.

\bibitem[Jevremovic et~al., 2018]{jevremovic2018wide}
Jevremovic, A., Ristic, N., Veinovic, M., Shimic, G., and Stanisic, N. (2018).
\newblock Wide: Centralized and collaborative approach to teaching web
  development.
\newblock {\em Journal of Internet Technology}, 19(4):1003--1014.

\bibitem[KEMP, 2022]{digital22}
KEMP, S. (2022).
\newblock Digital 2022 global overview report.
\newblock Technical report, DATAREPORRER.

\bibitem[Lundberg et~al., 2021]{lundberg2021employable}
Lundberg, G.~M., Krogstie, B.~R., and Krogstie, J. (2021).
\newblock From employable to fully operational: The need for training of
  computer science graduates.

\bibitem[Motogna et~al., 2021]{motogna21}
Motogna, S., Suciu, D., and Molnar, A.-J. (2021).
\newblock Investigating student insight in software engineering team projects.
\newblock In {\em Proceedings of the 16th International Conference on
  Evaluation of Novel Approaches to Software Engineering - ENASE,}, pages
  362--371. INSTICC, SciTePress.

\bibitem[Park and Wiedenbeck, 2011]{park2011learning}
Park, T.~H. and Wiedenbeck, S. (2011).
\newblock Learning web development: Challenges at an earlier stage of computing
  education.
\newblock In {\em Proceedings of the seventh international workshop on
  Computing education research}, pages 125--132.

\bibitem[Petrescu and Motogna, 2023]{petrescu23}
Petrescu, M. and Motogna, S. (2023).
\newblock A perspective from large vs small companies adoption of agile
  methodologies.
\newblock In {\em Proceedings of the 18th International Conference on
  Evaluation of Novel Approaches to Software Engineering}, pages 265--272.
  INSTICC, SciTePress.

\bibitem[Pop and Petrescu, 2023]{pop23}
Pop, E. and Petrescu, M. (2023).
\newblock Student's attraction for a carrier path related to databases and sql:
  Usability vs efficiency in students' perception -case study.
\newblock pages 182--189.

\bibitem[Radermacher et~al., 2014]{radermacher2014investigating}
Radermacher, A., Walia, G., and Knudson, D. (2014).
\newblock Investigating the skill gap between graduating students and industry
  expectations.
\newblock In {\em Companion Proceedings of the 36th international conference on
  software engineering}, pages 291--300.

\bibitem[Ralph, 2021]{ACM}
Ralph, P. (2021).
\newblock {ACM Sigsoft Empirical Standards for Software Engineering Research},
  version 0.2.0.

\bibitem[Redmond et~al., 2013]{redmond13}
Redmond, K., Evans, S., and Sahami, M. (2013).
\newblock A large-scale quantitative study of women in computer science at
  stanford university.
\newblock In {\em Proceeding of the 44th ACM technical symposium on Computer
  science education}, pages 439--444.

\bibitem[Statista, 2022]{Statista}
Statista (2022).
\newblock Data volume of global consumer internet traffic from 2017 to 2022, by
  subsegment.
\newblock Technical report, Statista.

\end{thebibliography}
{\small

\end{document}